\documentclass[11pt]{article}

\usepackage[preprint]{acl}
\usepackage{times}
\usepackage{listings}
\usepackage{latexsym}
\usepackage{amsmath}
\usepackage{float}
\usepackage[T1]{fontenc}

\usepackage[utf8]{inputenc}

\usepackage{microtype}

\usepackage{inconsolata}

\usepackage{graphicx}

\title{\includegraphics[width=1.5cm]{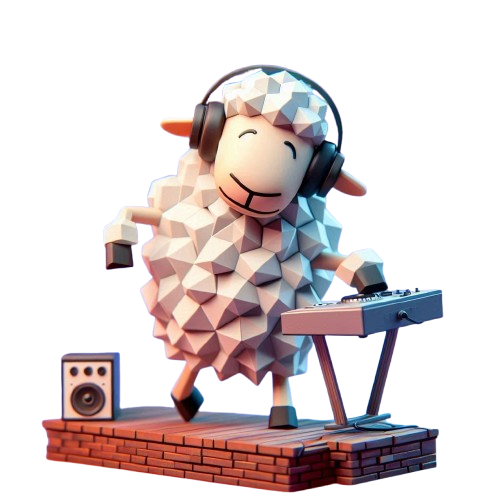}Sing-On-Your-Beat: Simple Text-Controllable Accompaniment Generations}

\author{Quoc-Huy Trinh$^{1,2,3}$, Minh-Van Nguyen$^{1,3,4}$, Trong-Hieu Nguyen Mau $^{1,4}$ \\ {\bf Khoa Tran } $^{1}$ \and {\bf Thanh Do} $^{1}$ \\
        $^{1}$SongGen Team, Ho Chi Minh city, Vietnam \\ 
        $^{2}$Aalto University, Espoo, Finland \\
        $^{3}$Technical University of Denmark, Kongens Lyngby, Denmark \\
        $^{4}$University of Science, VNU-HCM, Vietnam}

\begin{document}
\maketitle
\begin{abstract}
Singing is one of the most cherished forms of human entertainment. However, creating a beautiful song requires an accompaniment that complements the vocals and aligns well with the song's instruments and genre. With advancements in deep learning, previous research has focused on generating suitable accompaniments but often lacks precise alignment with the desired instrumentation and genre. To address this, we propose a straightforward method that enables control over the accompaniment through text prompts, allowing the generation of music that not only complements the vocals but also aligns with the song’s instrumental and genre requirements. Through extensive experiments, we successfully generate 10-second accompaniments using vocal input and text control. Additionally, our method demonstrates robust control over the generated accompaniment based on input prompts, improving alignment with the song’s instrumental and genre needs. A link to our work is available at https://songgen-ai.github.io/llambada-demo/.

\end{abstract}

\section{Introduction}
Music is a fundamental aspect of human culture. In recent years, with the rapid advancement of deep learning, several methods for non-vocal music generation have been proposed, such as AudioLM \cite{audiolm}, MusicGen \cite{musicgen}, and Stable Audio \cite{stableaudio}. These methods have shown promising results in addressing music generation challenges. In addition to non-vocal music, vocal music generation has become a significant task today, with various real-world applications. However, generating vocal music requires both the singing voice and instrumental accompaniment, as they play crucial roles in the process.

Naturally, in singing, the vocals and melody align with the beat, essential to forming a complete song. For this reason, SingSong \cite{singsong} and Fast-SAG \cite{fastsag} are two of the first systems designed to generate instrumental audio that accompanies vocal recordings. Regarding SingSong, they propose a system that adapts AudioLM \cite{audiolm} with the Transformer model to train an "audio-to-audio" model for the accompaniment generation via the vocal input. With FastSAG, they propose a Non-Autoregressive diffusion-based framework that is conditioned based on the vocal signal to directly generate the Mel-Spectrogram of the accompaniment. From their published results, the quality of the output audio is superior and their results are promising and have the potential to enhance music production.

Since the beat and accompaniment reflect the creativity of the producer, users often want \textbf{control} over these elements, including the specific instruments, beat type, rhythm, and tempo. Despite the promising audio quality of previous works, they have struggled to generate accompaniment that meets users' creative demands. To address this challenge, we propose \textbf{Llambada}, a simple, text-controllable system for accompaniment generation. Llambada allows users to define their desired accompaniment by inputting their requirements through a text prompt, enabling the input vocal sings along to their custom beats. This model incorporates AudioLM \cite{audiolm}, the Audio-Language Alignment Model (CLAP) \cite{elizalde2023clap}, Encodec \cite{Encodec}, and Mert \cite{mert}. These components process text prompt tokens alongside vocal semantic tokens—representing the general rhythm and harmony of the accompaniment—to generate accompaniment that is both in sync with the vocal input and text-controllable. To our knowledge, our system is one of the first systems that support the accompaniment generation with text control.

Moreover, the challenge of this task is the \textbf{lack of a dataset}. Previous works focus on the dataset with text for non-vocal music, and they do not concentrate on the prompt for the full song or the accompaniment. Due to the challenge of the dataset, we also propose the pipeline to generate the vocal and accompaniment instrumental with text prompt control to solve the challenge of text-controllable music generation, which can facilitate the dataset creation for the training of the model, thus encouraging the further research to explore and improve the quality of the accompaniment generation tasks. In our work, through extensive experiments, we successfully generate the 10-second accompaniment from vocal input with text control, which illustrates the promise of our method.

\begin{itemize}
    \item We propose a system that generates the accompaniment from vocal input with the control from the text prompt that can allow the user to control the accompaniment generation via prompt input.
    \item We propose a dataset pipeline to support the accompaniment via text control tasks, which can be further used for other tasks of music generation.
    \item We conduct extensive experiments to evaluate the effectiveness of the Llambada models. Experimental results show that our Llambada model can generate high-quality music, with text effectively controlling the instrumentation and song type. Additionally, we successfully achieve 10-second accompaniment generation using vocal input and text control.
    \item We open-source the implementation and the pre-trained (if possible) of the Llambada for further research in the accompaniment generation. 
\end{itemize}
This paper is organized as follows: in Section~\ref{sec:relatedwork} we briefly review existing methods related to this research. Then we propose our methods in Section~\ref{sec:method}. Experiments setup are in Section~\ref{sec:exp}. Finally, we present the conclusion in Section~\ref{sec:conclusion}.
\section{Related work}
\label{sec:relatedwork}
\subsection{Music Generation}
Audio generation is one of the most challenging tasks in AI. It involves generating audio from various inputs, such as speaker identity or reference speakers. One of the pioneering works in this field is WaveNet \cite{van2016wavenet}, which introduced a groundbreaking unconditional model to generate audio. WaveNet has been widely applied in text-to-speech systems. In the domain of music generation, it has also been used to produce piano and general music audio. Building on this, SampleRNN \cite{samplernn} was proposed, which utilizes RNNs at different scales of the model to capture long-term dependencies, addressing the limitations of previous works that struggled with maintaining coherence and rhythm in the generated audio.

In recent years, the rapid development of attention mechanisms and Transformer models \cite{vaswani2017attention} for sequence-to-sequence tasks has significantly advanced audio generation. One of the most notable contributions is Jukebox \cite{dhariwal2020jukebox}, which employs a VQ-VAE architecture and an autoregressive Transformer (AR) to generate high-fidelity music. A key innovation of Jukebox is its conditioning mechanism, which allows for input prompts such as vocal style and unaligned lyrics, resulting in highly coherent vocal performances and songs. In addition to Jukebox, recent models like AudioLM \cite{audiolm}, MusicLM \cite{agostinelli2023musiclm}, and MusicGen \cite{musicgen} have also gained popularity for generating high-quality music based solely on text prompts. These models introduced the concepts of semantic and coarse tokens, which have laid the groundwork for subsequent advancements in music generation, such as SingSong \cite{singsong}, and have influenced tools for music production, including SUNO and Udio.

Inspired by these works and their proven effectiveness, Llambada is designed with two stages. The first stage, semantic generation, produces semantic tokens that represent the general structure and rhythm of the song, controlled by vocal and prompt inputs. The second stage, coarse generation, produces acoustic tokens, which are decoded into music audio using the Encodec model \cite{Encodec}.
\subsection{Accompaniment Generation}
Accompaniment generation is the task of creating instrumental music that complements input vocals. SingSong \cite{singsong} and FastSAG \cite{fastsag} are among the first methods to successfully generate high-fidelity instrumental audio without relying on symbolic representations. Their results have been widely praised for their high-quality music output. However, a major challenge in previous work is the \textbf{lack of control} over the generated accompaniment, which can result in music that does not meet user expectations, making it difficult to use in music production.

To address this challenge, text-prompt control becomes crucial, as it serves as the interface through which users can interact with the model and guide the generated accompaniment to meet their desired outcome. In response to this need, Llambada is developed to allow users greater control and interaction during the accompaniment generation process.
\section{Task Definition and Method}
\label{sec:method}
\subsection{Task Definition}
\label{subsec:task_dev}
In this work, we propose an accompaniment generation model as a conditional model, with the inputs are vocal waveform $x_{vocal} \in R^{f_{s}T}$, and text prompt $X_{prompt}$ and the output is the accompaniment music waveform $Y \in R^{f_{s}T}$, with given sample rate $s$ and time frame $T$, we want to model a distribution $P(Y|X_{vocal}, X_{prompt})$. The effectiveness of the text control is assessed in Section~\ref{sec:exp}.

\subsection{Modelling through a proxy distribution}
As observed in previous works like Jukebox and WaveNet, modeling the distribution directly in the waveform domain requires significant computational resources and large datasets to generate high-quality audio, due to the complex structure of waveforms. To address this challenge, modeling the distribution in a proxy domain—using discrete audio codes—has proven more efficient. This approach simplifies training by constraining the problem to a finite set of audio codes. Specifically, we define the encoded vocal and accompaniment as $ \hat{X}{vocal} = \text{Enc}{vocal}(X)$ and $\hat{Y} = \text{Enc}(Y)$, where $\hat{X}{vocal} \in V^{f_{c}{T}}$ and $\hat{Y} \in V^{f_{c}{T}}$ (with $V$ is the finite vocabulary set with size $c$). These encoded representations can be decoded back into $X$ and $Y$ through the $\text{Dec}(.)$ function, where both the encoder and decoder are parametric functions. In terms of prompt input, we also do the transform from the text prompt embedding to the same discrete code domain of the audio, the audio code denotes as $\hat{X}_{prompt} \in V^{f_{c}{T}}$.


\subsection{Audio and Text representation}

Inspired by the MusicLM \cite{agostinelli2023musiclm}, we extract the semantic, and acoustic information of the audio by two public models are MERT \cite{mert}, and Encodec \cite{Encodec}. Then, through quantization, we transform these embeddings into discrete code tokens for the approximation via the proxy distribution. For text embedding representation, we leverage CLAP \cite{elizalde2023clap}, which is a pre-trained model to align text with audio data.


\textbf{Encodec} In this work, we use Encodec \cite{Encodec}, which is proposed to generate a high-quality codec representation of the audio for several audio downstream tasks. From the input audio waveform $X$, we extract the  75 Hz acoustic embedding. Then through the quantization with 12 quantizers and the codebook size of 1024, it results in the token sequence with a bitrate of 6kps, where 75 tokens are used to present 1 second of the audio, which denotes $Coarse(X)$. These acoustic tokens contain the audio quality, and the vocal identity of the audio. This information is leveraged in our method to support the model in understanding the quality of the input vocal, and also the speaker identity (which is empirically proved in the SingSong \cite{singsong}), which can support the generated accompaniment with more alignment between the vocal and the accompaniment. 

\textbf{Mert:} Similar to the concept of MusicLM \cite{agostinelli2023musiclm}, we use the immediate layers from the masked-language-modeling module of the MERT model \cite{mert} (95M parameters), which is the state-of-the-art model in the music understanding to extract the semantic embedding of the audio waveform input $X$. Then this embedding is quantized by using the K-means clustering with 1024 clusters over the embedding with the sampling rate of the input audio at 50 Hz. The codebook size of this token sequence is 1024. As the assumption from SingSong \cite{singsong}, the semantic tokens denote $Sem(X)$, and contain information of the rhythm, structure, and coherence of a music segment. The reason why we employ these semantic tokens is due to their ability to control the rhythm, content, and structure of the accompaniment (mentioned in AudioLM \cite{audiolm}) which is crucial information to form a song that has structure.

\textbf{CLAP:} To generate text representation tokens $CLAP(X_{prompt})$, we use CLAP \cite{elizalde2023clap} to extract text embeddings from the input prompt. A key advantage of the CLAP model is its robust training on aligning text with music and audio data, which enhances alignment with both the semantic and coarse acoustic embeddings, improving the overall robustness of the model. To transform these embeddings into a discrete code distribution (with a codebook size of 1024), we apply quantization using 12 quantizers.

Unlike MusicLM \cite{agostinelli2023musiclm}, we observed that training the model on audio data but performing inference on text can cause misinterpretations, resulting in noisy output. To address this issue, we propose a pipeline that generates pseudo prompts during training, allowing the model to train with real text prompts and infer with new ones, improving the consistency of the generated output.


\subsection{Overview of Llambada}
\begin{figure}[ht]
    \centering
    \includegraphics[width=1\linewidth]{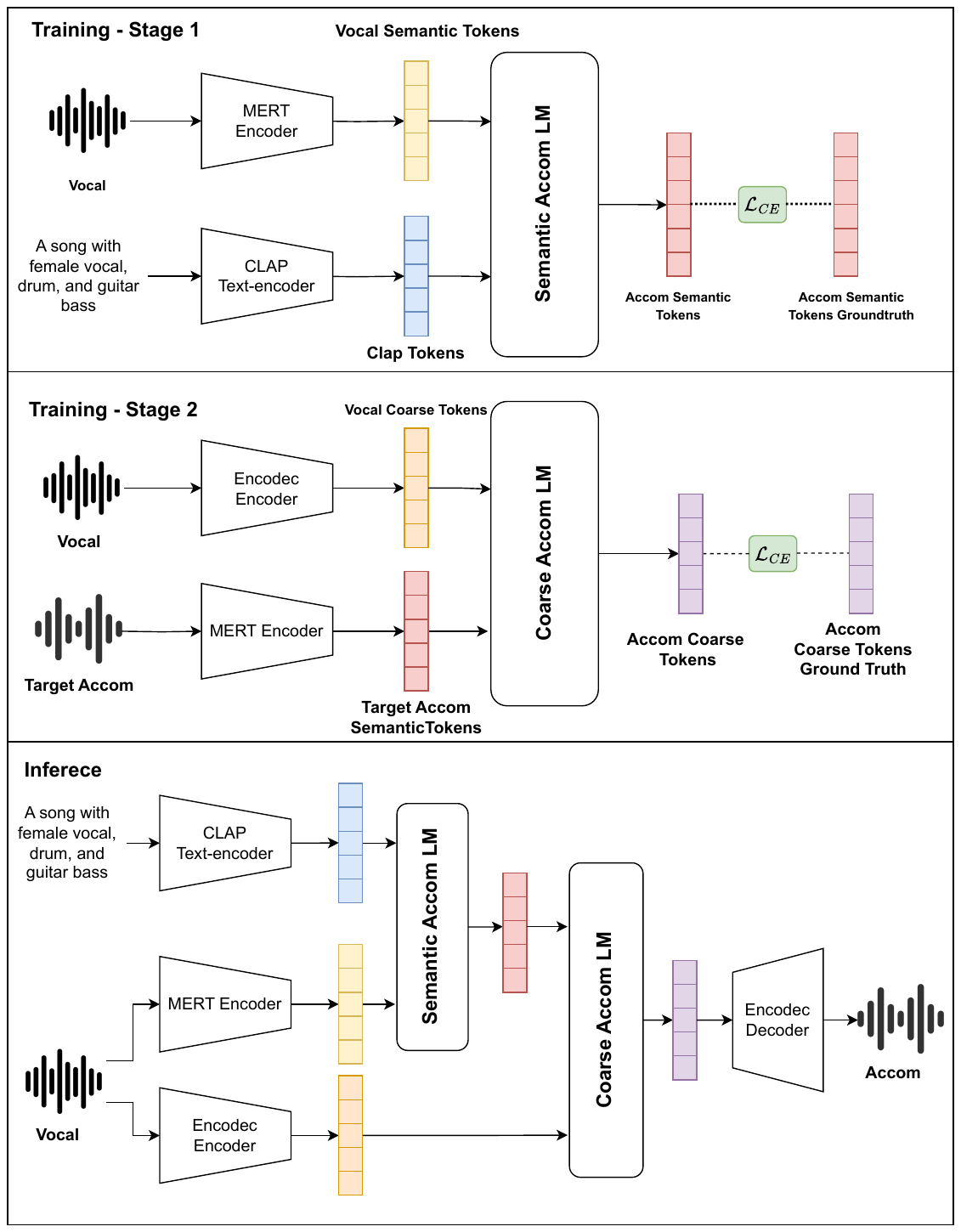}
    \caption{Visualization of 2 stages during training and inference of the Llambada model. Each token type is in the color as following: \textcolor{yellow}{vocal semantic tokens}, \textcolor{orange}{vocal coarse tokens}, \textcolor{blue}{clap text token}, \textcolor{red}{accom semantic token}, and \textcolor{purple}{accom coarse token}}
    \label{fig:vachi}
    \vspace{-5mm}
\end{figure}

As mentioned in the previous sections for the reason why we should do the modeling through a proxy distribution, Llambada is proposed to generate the acoustic tokens that can be decoded to the accompaniment waveform, which is illustrated in Figure~\ref{fig:vachi}. The first stage is the semantic modeling (input: vocal and prompt, output: semantic tokens of the accompaniment), and the Coarse Acoustic Modeling (input: accompaniment semantic tokens and the vocal coarse tokens). Let $X_{vocal} = A$, and text prompt $X_{prompt} = B$. From the task definition, we can decompose the modeling of the proxy distribution as the following Equation~\ref{equa:1}.

\begin{equation}
\centering
\small
\begin{split}
        P(Y|A, B) \approx P(\hat{Y} | \hat{A}, \hat{B}) \\
        = P(\text{Coarse}(Y)|\text{Sem}(Y), \text{Coarse}(A))\\.P(\text{Sem}(Y)|\text{Sem}(A), \text{CLAP}(B))
\label{equa:1}
\end{split}
\end{equation}

\noindent From the $\hat{Y}$, we can employ the Encodec \cite{Encodec} decoder model to decode the acoustic to the waveform which is the accompaniment output.


\subsection{Accompaniment Semantic generation stage}

As shown in Figure~\ref{fig:vachi}, during the accompaniment semantic generation stage, the vocal mono-audio waveform with dimensions $T \times 1$ (where $T$ represents the number of audio time frames) is encoded into an audio embedding using the MERT model \cite{mert}, resulting in an embedding with dimensions $T_{c} \times 768$ (where $T_{c}$ is the MERT timestep, depending on the audio input length). Through a quantization process, a code sequence with dimensions $T_{c} \times 5$ is obtained. Similarly, the text prompt is encoded by the CLAP model \cite{elizalde2023clap}, producing an embedding with dimensions $1 \times 512$, which is then quantized to a shape of $1 \times 12$, where 12 represents the number of quantizers. Finally, a T5 Transformer Decoder \cite{raffel2020exploring} is used to decode the accompaniment semantic tokens with dimensions $T_{c} \times 5$, representing the accompaniment and capturing structural, rhythmic, and latent features such as genre and instrumentation.


\subsection{Accompaniment Acoustic generation stage}

In the generation of accompaniment acoustic tokens, the input vocal mono-audio waveform with shape $T \times 1$ is processed by the Encodec model \cite{Encodec}. These are then encoded into a codec token sequence with shape $T_{e} \times n_{q}$, where $n_{q}$ is the number of quantizers, and $T_{e}$ represents the time steps of the Encodec model. This token sequence is integrated with the accompaniment semantic tokens of shape $T_{c} \times 1024$, and a T5 Transformer Decoder is employed to predict coarse acoustic tokens with shape $T_{e} \times 3$. These predicted tokens can then be decoded into the accompaniment waveform, maintaining the same shape as the vocal input.

\subsection{Llambada Inference Stage}
During the inference stage, users can input the text prompt and the vocal input for the accompaniment generation model. Then the CLAP model, MERT encoder, and Encodec model generate the code tokens for the prompt input, the semantic tokens, and the acoustic tokens of the vocal. In the first stage, the Semantic Accompaniment Language model uses the clap tokens and semantic tokens to predict the accompaniment semantic tokens. Then, in the next stage, the predicted accompaniment semantic tokens are inferred to the acoustic tokens of the accompaniment, which are then decoded to generate the final accompaniment.

\subsection{Pseudo Captioning dataset pipeline}

The community has released several music captioning models such as K2C-Aug \cite{K2c_aug}, and LP-MusicCaps \cite{musiccap} with promising results. Moreover, the main problem of the song dataset is that, there are no captions and descriptions for the song, and the annotation cost by humans is costly. For this reason, we leverage LP-MusicCaps for prompt generation. Get the input audio $X_a$, our target prompt $X_p$ is generated by the captioning model. Then, to filter out the unused data, and keep the important tags, we use Llama 3 \cite{llama3modelcard} model to do the tag extracting. From then, we can have clean prompts for the training dataset. The prompt output has the same format as the normal person's prompt to the system.

\section{Experiments}
\label{sec:exp}
\subsection{Dataset}
For the training and testing dataset, we all separate the source music into the two components are the vocal and the accompaniment through the state-of-the-art model in the music source separation, Demucs \cite{htdemuc1, htdemuc2}. Then, to assess the performance of the Llambada model by reproducing the training and testing Llambada with previous methods are SingSong and Fast-SAG in the following dataset setup.
\subsubsection{Training dataset}
For the training dataset, we do the training on the 4400 music hours for both vocals and the accompaniment, including the music some several genres such as pop, rock, jazz, and ballad and with the variant of the instruments including drum, bass, guitar, piano, and organ. For the prompt of the audio, we follow the format that includes genre, instruments, and the rhythm of the song, which can allow the model to learn and can generate the song with the following demands during the inference stage. Note that, all of the data are segmented into several 10s segments.
\begin{table*}[ht]
    \centering
    \begin{tabular}{|c|c|c|c|c|c|}
    \hline
      Method & $\text{FAD}_\text{VGGish}$ $\downarrow$ & $\text{FAD}_\text{Clap-music}$ $\downarrow$ &$\text{FAD}_\text{mean}$ $\downarrow$ & CLAP $\uparrow$\\
      \hline
      SingSong & 3.851 & 0.735& 2.293 & 0.157 \\
      FastSAG & 7.187 & 1.369 & 4.278 & 0.209 \\
      \hline
      Llambada & \textbf{3.156} & \textbf{0.679} & \textbf{1.918} & \textbf{0.244}\\
      \hline
    \end{tabular}
    \caption{In-domain testing }
    \label{tab:in_dist}
\end{table*}

\begin{table*}[ht]
    \centering
    \begin{tabular}{|c|c|c|c|c|c|}
    \hline
      Method & $\text{FAD}_\text{VGGish}$ $\downarrow$ &  $\text{FAD}_\text{Clap-music}$ $\downarrow$ & $\text{FAD}_\text{mean}$ $\downarrow$ & CLAP $\uparrow$ \\
      \hline
      SingSong & 6.650 & 0.647 & 3.649  & 0.166 \\
      FastSAG & 7.336 & 0.935 & 4.134& 0.148 \\
      \hline
      Llambada & \textbf{3.762} & \textbf{0.482} & \textbf{2.122} & \textbf{0.280} \\
      \hline
    \end{tabular}
    \caption{Out-of-distribution testing }
    \label{tab:out_dist}
\end{table*}

\subsubsection{Test dataset}
For the testing dataset, we follow the baseline of the creation of the training dataset, and we have two domains of testing.\\
\noindent \textbf{In-domain testing:} which uses the vocal and prompt from the same genre, with similar instruments from the training dataset (but these songs are not used in training) to evaluate the alignment and the effectiveness and the accompanied of the beat generated by the Llambada model in the accompaniment generation stage. This dataset setup contains 53 songs from a variety of instruments such as guitar, bass, and drum, and several genres such as pop, rock, ballad, and romantic. Each song is segmented into several 10-second segments for the best of comparison. \\

\noindent \textbf{Out-of-distribution testing} In the out-of-distribution test dataset setting, we reproduce the caption generation and preprocessing by the Musiccap \cite{agostinelli2023musiclm} for the benchmark matching between the output audio and the prompt. With the out-of-distribution dataset, we use the dataset from MusDB18 \cite{musdb18}, which is not included in our training dataset. For this testing dataset, we used 150 songs from the original set. This test dataset setting aims to evaluate the generalization and the robustness of the model.  

\noindent \textbf{Evaluation:} To evaluate audio quality and the alignment between the prompt and audio output, we use the Frechet Audio Distance (FAD) metric \cite{kilgour2018fr} and the CLAP score \cite{elizalde2023clap}. The FAD metric does not always accurately reflect audio output quality, so we reproduce FAD using two pre-trained models: VGGish \cite{hershey2017cnn} and CLAP music \cite{elizalde2023clap}. We then calculate the mean of these two results to enhance the robustness of the comparison.

\subsection{Baseline and Implementation Details}
As there are no official implementation and dataset details of the SingSong and FastSAG, we reproduce these two methods in as same as our dataset format for a fair comparison.

\textbf{SingSong (2023):} \cite{singsong} In our implementation of SingSong, we build upon the open-musiclm source code. Similar to MusicLM \cite{agostinelli2023musiclm} and the Llambada model, we utilize Encodec \cite{Encodec} and MERT \cite{mert}, the latter serving as a replacement for w2v-BERT \cite{hsu2021hubert}. During the implementation, we trained the model in two distinct stages: the semantic stage and the coarse acoustic stage. The semantic stage was trained for 200k steps and the coarse acoustic stage for 150k steps, each using a single A100 GPU with 80GB VRAM over approximately 5 and 6 days, respectively. The batch size and accumulation steps for the semantic stage were set to (32, 2), and for the coarse acoustic stage, they were (16, 8).

\textbf{FastSAG (2024):} \cite{fastsag} We reproduced the experiment on the public repository of the FastSAG and followed their setting, the training time cost us about 2 days on 1 A100 with 80GB GPU.

\textbf{Llambada:} We trained our Llambada model on two distinct stages are semantic stage and on two GPUs NVIDIA A100 80GB with about 100k steps with the learning rate of $1e-4$ and using the AdamW optimizer. The batch size and the accumulation step is set as (32, 4) and (16,8) for the semantic stage and coarse stage. Our training costs about 5 days for the semantic stage, and 4 days for the coarse stage.

\subsection{Qualitative results}
\subsubsection{In-domain results}
Table~\ref{tab:in_dist} presents a comparison of the Llambada model with previous works. Thanks to the text-prompt control, Llambada outperforms both SingSong and other Llambada variants across all four metrics, with particularly notable improvements in the CLAP score and $FAD_{mean}$, by $+0.087$ and $-0.0695$, respectively. These results indicate higher audio quality and stronger alignment between the text prompts and the generated music, making our model highly promising for text-based controllable music generation.
\subsubsection{Out-of-distribution results}

Table~\ref{tab:out_dist} illustrates the comparison between the Llambada model and previous methods on the out-of-distribution testing domain. Thanks to text-prompt control, our method surpasses prior works on the $FAD_{mean}$ and $CLAP$ score metrics, with improvements of $-1.537$ and $+0.12$, respectively. These results demonstrate the significant enhancement in Llambada's performance when guided by text prompts, which align the model’s output more closely with the ground truth, leading to better benchmark results.

\subsection{Ablation Study}
To assess the effectiveness of the text guiding in several criteria from the song, we do the experiments to validate the matching between the prompt control with each criteria of the music, including the instruments and the genres of the song. However, to do that we have to have the music expert do the evaluation, but it is difficult, and high cost for the music expert's judgment. Thanks to the Llava-Bench \cite{liu2024visual} idea when they leverage the GPT-4 in the benchmark of the multimodal-LLM, inspired by this idea, we employ Qwen-Audio \cite{chu2023qwen}, an Audio-Language model to the judgment and scoring between the alignment of the music generated output and the ground truth in several music criteria. The benchmark based on Qwen-Audio is produced three times, and we calculate the mean score of the three-time outputs. The example of the templates to ask the grading from Qwen-Audio is the following:

\begin{enumerate}
    \item Human: Given the <audio input>, with prompt <prompt value>, as a producer, can you give the score for the alignment between the <genre/instruments> of the song with prompt input on a scale from 0 to 100?
    \item Qwen-Audio: As a producer the score in the alignment between the song and the prompt in <genre/instruments> is <score>.
\end{enumerate}

From then, we extract the score from the results and store them until we get 3 responses for the mean calculation. All of the studies are conducted with the in-domain dataset.

\subsubsection{Evaluation of the effectiveness of the text prompt in the instrument alignment}

To assess the effectiveness in the alignment between the text prompt and the instrumental from the music output, we do the benchmark on the instrumental characteristic criteria of the song with the instrument mentioned in the accompaniment ground truth, which is illustrated in Table~\ref{tab:inst}. 
\begin{table}[H]
    \centering
    \begin{tabular}{|c|c|}
    \hline
        \textbf{Baseline} & \textbf{Alignment score} \\
        \hline
         \textbf{W/o} text control & $80.170  $ \\
         \textbf{With} text control & $82.950  $ \\
         \hline
     \end{tabular}
    \caption{Table shows the Alignment scores to assess the effectiveness of text guiding in the alignment of the instruments}
    \label{tab:inst}
\end{table}

From the results, we observe that with the text prompt control, the alignment in the instrument performs better than the approach without text control. This result indicates the importance of prompts in guiding the accompaniment that can follow the instrumental demand of the client.

\subsubsection{Evaluation of the effectiveness of the text prompt in the genre alignment}
To evaluate the alignment in genre between the generated accompaniment that is controlled by text prompt and not controlled by text prompt, we do the evaluation based on the genre criteria, which is illustrated in Table~\ref{tab:genre}.
\begin{table}[H]
    \centering
    \begin{tabular}{|c|c|}
    \hline
        \textbf{Baseline} & \textbf{Alignment score} \\
        \hline
         \textbf{W/o} text control & $79.101  $ \\
         \textbf{With} text control & $80.415$ \\
         \hline
     \end{tabular}
    \caption{Table shows the relative scores to assess the effectiveness of text guiding in the alignment of the instruments}
    \label{tab:genre}
\end{table}

As can be seen, the alignment in the genre score of the model with the text control performs better than the model without text control, which demonstrates that the prompt can support the model in understanding the context of the audio, thus making the better in the genre alignment.

\section{Conclusion}
\label{sec:conclusion}
In conclusion, we introduce Llambada, a novel pipeline for the text prompt control of the accompaniment generation task. Additionally, we also propose a pseudio captioning dataset pipeline of the dataset creation for the text prompt from the accompaniment, which can serve this task. From extensive experimental results, it shows that our method can surpass in the music quality and the better alignment with the clients' demands when compare with previous works, which indicate the promising results of the model.

Although the promising performance of our method, there are several rooms should be improved. Firstly, the alignment between the text prompt with the semantic feature of the song to make the better understanding of the Language Model in the relation between inputs tokens. Secondly, this method leverages two Language models, which can get the extensive computing resource. We encourage the further researches focus on these two limitations to improve the model, thus can enhance the application of the music generation.

\bibliography{custom}



\end{document}